\documentclass[fleqn,usenatbib]{mnras}
\usepackage{hyperref}
\usepackage{newtxtext,newtxmath}
\usepackage{booktabs}

\usepackage[T1]{fontenc}
\usepackage{xcolor}

\DeclareRobustCommand{\VAN}[3]{#2}
\let\VANthebibliography\thebibliography
\def\thebibliography{\DeclareRobustCommand{\VAN}[3]{##3}\VANthebibliography}

\usepackage{graphicx}	
\usepackage{amsmath}	
\usepackage{float}
\usepackage{accents}
\usepackage{cancel}

\title[Modified gravity with cluster-lensed GWs]{Probing modified gravity with galaxy-cluster-lensed gravitational-wave time delays: impact of systematic effects}

\author[Tsaprazi et al.]{Eleni Tsaprazi$^{1,2}$ \thanks{e.tsaprazi@imperial.ac.uk, eleni.tsaprazi@obspm.fr}, Ettore Delpogetto$^{3}$, Federico Marulli$^{3,4,5}$, Amandine~M.~C.~Le~Brun$^{2}$
\\
$^{1}$Imperial Centre for Inference and Cosmology (ICIC) \& Astrophysics group, Department of Physics, Imperial College, Blackett Laboratory,\\ Prince Consort Road, London SW7 2AZ, UK\\
$^{2}$LUX, Observatoire de Paris, Université PSL, Sorbonne Université, CNRS, 92190 Meudon, France\\
$^{3}$Dipartimento di Fisica e Astronomia “Augusto Righi” - Alma Mater Studiorum Universita di Bologna, via Piero Gobetti 93/2, I-40129 Bologna, Italy\\
$^{4}$INAF - Osservatorio di Astrofisica e Scienza dello Spazio di Bologna, via Piero Gobetti 93/3, I-40129, Bologna, Italy\\
$^{5}$INFN - Sezione di Bologna, viale Berti Pichat 6/2, I-40127 Bologna, Italy
}

\date{Accepted XXX. Received YYY; in original form ZZZ}

\pubyear{2026}

\begin{document}
\label{firstpage}
\pagerange{\pageref{firstpage}--\pageref{lastpage}}
\maketitle

\begin{abstract}
Strongly lensed gravitational waves provide exceptionally precise measurements of relative macroimage arrival times, making their time delays a promising time-domain probe of gravity on galaxy-cluster scales. This precision places stringent demands on the control of lensing systematic effects. We present a proof-of-concept pipeline for constraining a phenomenological modification, $D$, of the galaxy cluster Weyl potential using the macroimage time-delay span of a simulated sample of binary neutron star (BNS) sources and cluster lenses described by a Navarro--Frenk--White cluster halo and a singular isothermal sphere for the brightest cluster galaxy (BCG). Closure tests recover the injected general-relativistic (GR) limit. Adopting a fiducial 5 per cent precision on the reconstructed time-delay span, we assess controlled mis-specifications of the halo mass and concentration, BCG contribution, source and lens positions, redshifts, line-of-sight environment, projected ellipticity and secondary cluster-scale structure. Within the adopted gravity model, neglecting halo ellipticity, secondary cluster-scale structure, cluster mis-centring and line-of-sight perturbations can induce significant, strongly configuration-dependent biases in the inferred parameter $D$. Biased estimates of the cluster halo mass and concentration, as well as systematic BNS source-position errors at the level considered here, produce significant coherent shifts. By contrast, uncertainties in the BCG contribution and in the lens and source redshifts are comparatively subdominant. These results highlight the need for explicit treatment of the above uncertainties before cluster-lensed gravitational wave time delays can be used as precision tests of gravity.
\end{abstract}




\section{Introduction}\label{sec:intro}

Despite the remarkable success of general relativity (GR), the unknown physical nature of dark matter and dark energy, together with persistent tensions in cosmological data, motivates testing gravity across a wider range of astrophysical and cosmological regimes \citep{2019LRR....22....1I}. Gravitational waves (GWs) have opened new avenues for this programme through tests of their polarisation, propagation and waveform generation \citep[e.g.][]{2016PhRvL.116f1101B,2018PhRvL.120t1102A,2021PhRvD.103l2002A,6c61-fm1n}. The multimessenger observation of GW170817 provided an especially stringent constraint on the GW propagation speed \citep{2017PhRvL.119p1101A}, excluding broad classes of modified-gravity and dark-energy models that predict appreciable deviations from the speed of light \citep{2017PhRvL.119y1304E}. This result narrowed the space of viable theories and reinforced the importance of complementary tests that probe gravity through its effects on cosmological observables. The next generation of wide-field surveys and GW detectors will substantially expand this programme by providing both larger data sets and qualitatively new probes of gravity.

One such novel probe is provided by strongly lensed GWs. Depending on the GW frequency and the lens mass, lensing can appear either in the geometric-optics regime, where multiple images arrive with different magnifications and time delays, or in the wave-optics regime, where interference produces frequency-dependent amplitude and phase modulations \citep{2025RSPTA.38340134S}. For GWs from compact binaries observed by ground-based detectors and lensed by galaxies or galaxy clusters, the geometric-optics approximation is well justified, as wave-optics effects are negligible \citep{2023Univ....9..200G}.

In this work, we develop a forward-modelling framework to test phenomenological modifications of gravity using time delays from cluster-lensed GWs in the geometric-optics regime, and quantify how inaccuracies in the cluster mass distribution, source reconstruction and line-of-sight environment can bias the inferred gravity parameter and mimic departures from GR. Cluster-lensed compact-binary events are expected to become increasingly relevant in the era of third-generation GW detectors. Their greater sensitivity will extend the detectable compact-binary population to higher redshifts, thereby yielding progressively larger samples of strongly lensed events
\citep{2022MNRAS.515.1044M,2023PhRvD.107j3023A,
2024ApJ...977...64C,2024arXiv241112453O,
2024MNRAS.535..990M}. Galaxy clusters are particularly interesting lenses because they can produce several well-separated macroimages with delays ranging from days to years, while the relative arrival times of the corresponding GW signals can be measured with very high precision. In addition, many cluster lenses can be identified and characterised through electromagnetic observations, providing redshifts, image configurations and independent constraints on the lens and foreground mass distribution \citep{2020MNRAS.495.3727R}.

Binary black hole (BBH) mergers are promising sources to be considered because they can be detected to large distances, while lensing magnification further increases the accessible source population \citep{2020JCAP...03..050M,2021ApJ...921..154W}. However, BBHs are not generally expected to produce identifiable electromagnetic counterparts because they contain little baryonic matter available to power transient emission. Binary neutron star (BNS) and neutron star--black hole (NSBH) mergers, on the other hand, provide a complementary multimessenger channel \citep{2017ApJ...848L..12A}. Electromagnetic information can aid the association of repeated lensed signals, enable host and source-redshift identification, and improve the reconstruction of the lens--source configuration. In this proof-of-concept analysis, we restrict the mock source population to BNSs as a representative multimessenger population. This choice is not specific to the macroimage time-delay observable itself, which does not depend directly on the binary type. Irrespective of the source population, accurate cluster modelling remains essential, since residual uncertainties in the halo, central galaxy, line-of-sight environment and source position can bias the interpretation of the measured time delays and mimic a modification of the lensing potential. While we do not explicitly model counterpart detectability or host-identification failures, we assess their relevant downstream modelling uncertainties through perturbations of the source redshift and source-plane position.

Time delays from lensed GWs have several features that make them attractive for tests of gravity. First, the arrival-time differences between lensed GW images may in principle be measured far more precisely than the time delays of multiply imaged quasars or supernovae \citep{2018IAUS..338...98S}. Second, clusters probe massive nonlinear structures at cosmological distances, providing access to gravitational potentials on scales complementary to those tested by galaxy lenses and large-scale structure. Third, if the foreground cluster and source configuration can be identified or externally constrained, the measured GW time delays provide a direct time-domain observable of the lensing potential. However, most existing studies of modified gravity with lensed-GW time delays have focused on lower-mass lenses \citep[e.g.][]{2019ApJ...880...50Y,2021arXiv210609630C,2021PhRvD.104h4057F,2025PDU....4701795T,2025MNRAS.536.2212P,2026arXiv260309340M}, while the cluster-lensing case remains comparatively less explored \citep{2018IAUS..338...98S,2025PhRvD.112f3044V,2026PhRvD.113b4067C}.

\citet{2019ApJ...880...50Y} propose two approaches to testing modified gravity with the strong lensing of GWs. Their first approach compares the GW time delay predicted in GR and in a phenomenologically modified Weyl potential, using the lens mass scale inferred from stellar velocity-dispersion measurements. Its sensitivity is therefore limited mainly by uncertainties in the dynamical mass calibration. Their second approach exploits the analytic consistency relation between image positions and time delay for a GR singular-isothermal-sphere (SIS) lens, thereby eliminating the velocity-dispersion measurement, but at the cost of a much stronger dependence on the assumed lens profile, since departures from the fiducial model can mimic a violation of the relation. Our method is closer to the first strategy: we forward-model the modified Weyl potential through a specified lens model, propagate it to the image configuration and time delay, and explicitly quantify how modelling uncertainties can bias the inferred modified-gravity parameter.

In the present analysis, we construct controlled mock catalogues to run through the pipeline in order to isolate the impact of individual systematic effects. We retain the commonly adopted decomposition of the cluster potential into a Navarro--Frenk--White (NFW) halo and an SIS central-galaxy component \citep{2023PhRvD.108j3529T,2026RAA....26f2001C}. NFW models can reproduce the dominant radial mass profiles of clusters at radii greater than $50$ kpc \citep[see][and references therein]{2013ApJ...765...24N}, while the SIS provides a low-dimensional description of the approximately isothermal central potential of massive early-type galaxies \citep[see][and references therein]{2010ApJ...724..511A}. In what follows, we specifically examine departures from this baseline arising from unmodelled cluster ellipticity, secondary structure, and line-of-sight contributions, shifts to the true source and lens positions, as well as redshift uncertainties. Future work aimed at characterising lens-modelling systematics could use cluster populations extracted from hydrodynamical simulations, particularly since \citet{2025PhRvD.112f3044V} showed that cluster substructure can alter time delays of lensed GWs. A realistic detector forecast would additionally require modelling the observable GW population and selection function, including which lensed binaries and macroimages are detectable.

Significant effort has been invested into studying GW lensing, including its detection \citep{2020arXiv201012093K,2020MNRAS.492.1127M,2021MNRAS.506.5430J,10.1098/rsta.2024.0129}, modelling \citep{2019ApJ...880...50Y,2021MNRAS.503.3326C,2024PhRvD.110j4050T,2025PDU....4701795T}, forecasting and survey-design studies \citep{2018MNRAS.476.2220L,2021ApJ...921..154W,2026PhRvD.113d3552A}, as well as cosmological inference \citep{2024PhRvD.109h4064N,2026PhRvD.113l3535C}. As the statistical power of next-generation data increases, lens-modelling systematic effects will become increasingly important for robust tests of gravity. Here, we develop a proof-of-concept pipeline for a conditional sample of simulated cluster-lensed GW systems to quantify how uncertainties in the lens, source and environment can mimic a departure from GR. Our aim is to identify which modelling uncertainties can significantly bias modified-gravity inference and which therefore require explicit modelling, marginalisation, or external constraints, rather than to provide a full third-generation detector forecast. 

The paper is structured as follows. In Sec. \ref{sec:method} we describe the phenomenological modified gravity model we adopt, the mock lensed event generation, the likelihood and the systematic effects that we consider. In Sec. \ref{sec:results} we report our results on the impact of unmodelled systematic effects on the modified gravity constraints. In Sec. \ref{sec:conclusions} we summarise our findings and highlight advancements required on the theory and modelling front for GW lensing by galaxy clusters as a probe of gravity. 

\section{Method}\label{sec:method}

In this section we discuss the modified gravity model present in \citet{2019ApJ...880...50Y}, a variation of which we adopt in our proof-of-concept pipeline, we detail the mock data generation process, discuss our statistical framework and the interpolation scheme we implement to speed up the forward model evaluation, along with validation results.

\subsection{Modified lensing kernel}
\label{subsec:modified_lensing_kernel}

We consider departures from GR that act directly on the Weyl potential sourcing light deflection, while leaving the background geometry and the matter distribution fixed. We work in the conformal Newtonian gauge, for which the flat perturbed Friedmann--Lemaître--Robertson--Walker (FLRW) metric can be written as
\begin{equation}
\mathrm{d}s^2 =
\left(1+\frac{2\Psi}{c^2}\right)c^2 \mathrm{d}t^2
-
a^2(t)\left(1-\frac{2\Phi}{c^2}\right)\mathrm{d}\mathbf{x}^2 \, .
\end{equation}
Here $t$ is cosmic time, $a(t)$ is the scale factor, $c$ is the speed of light, $d\mathbf{x}$ is the comoving spatial line element, and $\Psi$ and $\Phi$ are the two scalar metric potentials. The Weyl potential, $\Phi_{+}\equiv (\Phi+\Psi)/2$, governs the gravitational lensing of electromagnetic radiation and GWs, and is therefore the metric-potential combination relevant to the lensing observable considered here. Following the phenomenological prescription of \citet{2019ApJ...880...50Y}, we parameterise modifications to the Weyl potential through a radial function $\Sigma(r)$, such that the point-mass Weyl potential can be written as
\begin{equation}
  \Phi_{+}(r)=-\frac{G M_{\rm lens}}{r}\Sigma(r)\, .
\end{equation}
Here $G$ is Newton's gravitational constant, $M_{\rm lens}$ is the lensing mass entering the point-mass kernel, $r$ is the three-dimensional physical distance from the lens centre, and $\Sigma(r)$ encodes deviations from the GR Weyl potential. In GR, $\Sigma(r)=1$.

We assume that the modification is screened below a transition scale $R_\mathrm{scr}$ and adopt the radial form
\begin{equation}
\Sigma(r;D,A,R_\mathrm{scr})=
\begin{cases}
1\,, & 0<r<R_\mathrm{scr}\,,\\[5pt]
\dfrac{
1+\left(A+\dfrac{D}{2}\right)(r-R_\mathrm{scr})^2
}{
1+\left(A-\dfrac{D}{2}\right)(r-R_\mathrm{scr})^2
}\,, & r\ge R_\mathrm{scr}\,,
\end{cases}
\label{eq:sigma_DAR}
\end{equation}
which is a reparameterised version of the model of \citet{2019ApJ...880...50Y}\footnote{In the notation of \citet{2019ApJ...880...50Y}, our parameters correspond to
$D=(\alpha_1-\alpha_2)/w_{01}^{2}$, $A=(\alpha_1+\alpha_2)/(2w_{01}^{2})$, and
$R_\mathrm{scr}=r_{01}$.}. The parameter $D$ controls the signed amplitude of the departure from GR outside this scale, while $A$ controls the radial shape of the numerator and denominator. The above formalism presented by \citet{2019ApJ...880...50Y} draws from earlier work by \citet{PhysRevD.78.024015,PhysRevD.86.123503,PhysRevD.94.104014}. Since $r$ and $R_\mathrm{scr}$ are physical lengths, both $D$ and $A$ have dimensions of inverse length squared. The GR limit is recovered for $D=0$. In the present proof-of-concept analysis, we fix $R_\mathrm{scr} = 1\,{\rm kpc}\,$ and $A = 1\,{\rm kpc}^{-2}$ and infer only the parameter $D$. With these choices, both the screening scale and the characteristic transition scale are well inside the cluster strong-lensing region considered here. The images therefore predominantly probe the asymptotic regime of the phenomenological modification, in which $D$ acts approximately as a rescaling of the effective Weyl lensing strength rather than as a strongly radius-dependent deformation.

The parameter $D$ is a phenomenological kernel parameter rather than a fundamental parameter of a particular theory of gravity. It controls the amplitude of the adopted radial deformation of the Weyl potential, but it does not correspond directly to a parameter appearing in a specific modified-gravity Lagrangian. This distinction is important: while \(D=0\) is the GR limit of our phenomenological kernel, a nonzero value of \(D\) cannot be identified directly with a specific theory of gravity. Such a mapping would require solving the chosen theory around a cluster-scale lens, including its screening mechanism, and projecting the resulting \(\Phi_+(r)/\Phi_+^{\rm GR}(r)\) profile onto our kernel. Our results should therefore be read as constraints on a screened Weyl-potential deformation motivated by modified-gravity phenomenology, not as direct constraints on named modified-gravity models. Therefore, the question we aim to address with this model is whether cluster-lensed GW time delays constrain or falsely detect $D\neq 0$ under modelling systematic effects. Future work should aim to connect this phenomenological framework to concrete modified-gravity models by deriving their nonlinear, screened cluster-lensing predictions and projecting the resulting Weyl-potential modifications onto observables such as image positions, magnifications, and time delays. If more than one of the parameters in $\Sigma(r;D,A,R_\mathrm{scr})$ is allowed to vary, strong degeneracies are expected. The time-delay data vector is sensitive to the integrated lensing potential evaluated at the image positions, rather than to the full radial form of $\Sigma(r)$. As a result, different combinations of $D$, $A$, and $R_\mathrm{scr}$ can produce very similar changes in the Fermat-potential differences. Breaking these degeneracies would require additional information beyond the single time-delay data vector used in this proof-of-concept analysis. 

For a thin lens, the time-delay difference between two lensed images of a transient GW event is
\begin{equation}
\Delta t_{ij}
=
\frac{1+z_\mathrm{l}}{c}\,
\frac{D_\mathrm{l}D_\mathrm{s}}{D_\mathrm{ls}}\,
\Delta\phi_{ij}\,,
\label{eq:time_delay}
\end{equation}
where $i$ and $j$ label the two images, $z_\mathrm{l}$ is the lens redshift, $D_\mathrm{l}$ is the angular-diameter distance from the observer to the lens, $D_\mathrm{s}$ is the angular-diameter distance from the observer to the source, and $D_{ls}$ is the angular-diameter distance between the lens and the source. The Fermat-potential difference is $
\Delta\phi_{ij}
\equiv
\phi(\boldsymbol{\theta}_i)
-
\phi(\boldsymbol{\theta}_j)$, where $\boldsymbol{\theta}_i$ and $\boldsymbol{\theta}_j$ are the angular positions of the two images. The Fermat potential is
\begin{equation}
\phi(\boldsymbol{\theta})
=
\frac{|\boldsymbol{\theta}-\boldsymbol{\beta}|^2}{2}
-
\psi(\boldsymbol{\theta})\,.
\end{equation}
Here $\boldsymbol{\theta}$ is an angular coordinate in the image plane, $\boldsymbol{\beta}$ is the unlensed angular source position, and $\psi(\boldsymbol{\theta})$ is the two-dimensional lensing potential, which is obtained by projecting the Weyl potential along the line of sight,
\begin{equation}
\psi(\boldsymbol{\theta})
=
\frac{2}{c^{2}}
\frac{D_{\rm ls}}{D_{\rm l}D_{\rm s}}
\int
{\rm d}z\,
\Phi_{+}\!\left(D_{\rm l}\boldsymbol{\theta},z\right),
\label{eq:lensing_potential}
\end{equation}
up to an irrelevant additive constant. Here, $z$ represents the physical coordinate along the line of sight through the lens and $c$ the speed of light. Since in this work the modification is applied only to the lensing potential, while the background cosmology is kept fixed, the angular-diameter distances retain their GR value.

For an extended lens, the modified Weyl potential enters the deflection angle through the projected mass distribution. Following \citet{2019ApJ...880...50Y}, in order to introduce the effect of modified gravity, we write the reduced deflection angle as
\begin{equation}
\boldsymbol{\alpha}(\boldsymbol{\xi})
=
\frac{4G}{c^{2}}\frac{D_\mathrm{ls}}{D_\mathrm{s}}
\int {\rm d}^{2}\xi'\,
\Theta(\boldsymbol{\xi}')
(\boldsymbol{\xi}-\boldsymbol{\xi}')
I\!\left(|\boldsymbol{\xi}-\boldsymbol{\xi}'|\right)\,,
\label{eq:angle_MG}
\end{equation}
where $\boldsymbol{\xi}=D_\mathrm{l}\boldsymbol{\theta}$ is the physical transverse coordinate in the lens plane, and $\boldsymbol{\xi}'$ is an integration coordinate in the lens plane. The projected surface mass density is
\begin{equation}
\Theta(\boldsymbol{\xi})
=
\int {\rm d}z\,\rho(\boldsymbol{\xi},z)\,,
\end{equation}
where $\rho(\boldsymbol{\xi},z)$ is the three-dimensional matter density of the lens. The modified radial kernel is
\begin{equation}
I(b)
=
\int_0^\infty {\rm d}z
\left[
\frac{\Sigma(r;D)}{r^3}
-
\frac{1}{r^2}\frac{d\Sigma(r;D)}{dr}
\right]_{r=\sqrt{z^2+b^2}}\,,
\label{eq:kernel_MG}
\end{equation}
where $
b=|\boldsymbol{\xi}-\boldsymbol{\xi}'|$ is the transverse separation between the field point and the mass element in the lens plane, and $
r=\sqrt{z^2+b^2}$ is the corresponding three-dimensional physical separation. In the GR limit, $\Sigma=1$ and $d\Sigma/dr=0$, so the standard lensing kernel is recovered. We omit dependence on $A, R_\mathrm{scr}$ above and in what follows for brevity. We implement this prescription directly in \texttt{lenstronomy} \citep{2018PDU....22..189B} by defining custom circular lens profiles. The cluster-scale dark-matter halo is described by an NFW component \citep{1997ApJ...490..493N} whose lensing potential is modified according to Eq. \eqref{eq:sigma_DAR}. We also include a central brightest cluster galaxy (BCG) contribution modelled as an SIS, to which we apply the same Weyl-potential modification. The lensing strength is parameterised by the Einstein radius \citep{1996astro.ph..6001N}
\begin{equation}
\theta_{\rm E,BCG}
=
4\pi
\left(\frac{\sigma_\mathrm{v}}{c}\right)^2
\frac{D_{ls}}{D_\mathrm{s}}\,,
\label{eq:SIS}
\end{equation}
where $\sigma_\mathrm{v}$ is the BCG velocity dispersion. The BCG velocity dispersion is assigned through a two-stage simulation-based prescription. We first infer the cluster velocity dispersion from the galaxy-tracer relation of \citet[][Table 1]{2013MNRAS.430.2638M} based on the results of simulations including feedback from active galactic nuclei for galaxy tracers, including an independent 12 per cent Gaussian scatter in linear velocity dispersion. We then obtain the BCG stellar velocity dispersion from the redshift-dependent relation of \citet[][Table 1]{2022ApJ...931...31S}. This prescription provides the fiducial BCG normalisation used for catalogue generation. 

The total lensing potential is therefore
$
\psi_{\rm tot}(\boldsymbol{\theta};D)
=
\psi_{\rm NFW}^{\rm Weyl}(\boldsymbol{\theta};D)
+
\psi_{\rm SIS}^{\rm Weyl}(\boldsymbol{\theta};D)$.
Here $\psi_{\rm NFW}^{\rm Weyl}$ is the modified lensing potential of the cluster-scale NFW halo and $\psi_{\rm SIS}^{\rm Weyl}$ is the modified lensing potential of the BCG-like SIS component. Given the total potential, the image positions are obtained by solving the lens equation
$
\boldsymbol{\beta}
=
\boldsymbol{\theta}
-
\boldsymbol{\alpha}(\boldsymbol{\theta};D)$,
where
$
\boldsymbol{\alpha}(\boldsymbol{\theta};D)
=
\nabla_{\boldsymbol{\theta}}\psi_{\rm tot}(\boldsymbol{\theta};D)$
is the angular deflection field. The image positions and the corresponding Fermat potentials are then used to compute the time delays through Eq. \eqref{eq:time_delay}. In the present analysis, we assume that the lensing cluster has been perfectly identified following \citet{2026RAA....26f2001C}.

\subsection{Mock-data generation}
\label{subsec}

We construct a fixed catalogue of 250 geometrically multiply-imaged galaxy-cluster--BNS lens--source configurations. The catalogue is generated once using a random seed, and subsequently held fixed throughout the baseline and systematic tests. It is conditional on the adopted theoretical halo and BNS populations, fiducial lens model, background cosmology and geometric multiple-imaging selection. It should therefore not be interpreted as an event-rate or detected-population forecast. We restrict the catalogue to BNSs as a representative multimessenger population, since an electromagnetic counterpart may enable host identification. Within the geometric-optics calculation adopted here, the macroimage time delays do not otherwise depend on the binary type. BBHs and NSBHs would differ primarily through their source-redshift and observational-selection distributions.

We first generate a proposal population of lens haloes. The lens redshift and logarithmic halo mass are drawn jointly from
\begin{equation}
p_{\rm l}(z_{\rm l},\ln M_\mathrm{200c})
\propto
\frac{{\rm d}V_{\rm c}}{{\rm d}z_{\rm l}{\rm d}\Omega}
\frac{{\rm d}n(M_\mathrm{200c},z_{\rm l})}{{\rm d}\ln M_\mathrm{200c}},
\label{eq:proposal}
\end{equation}
over $0.1<z_{\rm l}<1.0$ and $10^{14}<M_{200c}/M_{\odot}<10^{15}$. Here $M_\mathrm{200c}$ is the physical mass enclosed within $R_\mathrm{200c}$, where the mean density is 200 times the critical density at the halo redshift. We adopt the \citet{2008ApJ...688..709T} halo mass function and assign each halo the median \citet{2021MNRAS.506.4210I} concentration--mass relation using \texttt{COLOSSUS} \citep{2018ApJS..239...35D}. 

Source redshifts are drawn independently over $0.2<z_{\rm s}<5.0$ from the observer-frame distribution
\begin{equation}
p_{\rm s}(z_{\rm s})
\propto
\frac{{\rm d}V_{\rm c}}{{\rm d}z_{\rm s}{\rm d}\Omega}
\frac{R_{\rm BNS}(z_{\rm s})}{1+z_{\rm s}},
\label{eq:z_s}
\end{equation}
where $R_{\rm BNS}$ is the source-frame BNS merger-rate density and the factor $(1+z_{\rm s})^{-1}$ accounts for cosmological time dilation. We impose the physical foreground--background ordering $z_{\rm s}>z_{\rm l}$. The merger-rate history is obtained by convolving the cosmic star-formation-rate density with a delay-time distribution
\begin{equation}
R_{\rm BNS}(t)
\propto
\int_{t_{\min}}^{t_{\max}(t)}{\rm d}t_{\rm d}
\dot{\rho}_{\star}(t-t_{\rm d})
p(t_{\rm d}).
\label{eq:R_BNS}
\end{equation}
Here, $t$ is the cosmic time at merger, $t_{\rm d}$ is the delay
between binary formation and merger, and $t-t_{\rm d}$ is therefore
the corresponding formation time. The maximum allowed delay is
$t_{\max}(t)=t-t(z_{\rm form,max})$, where $t(z)$ denotes the cosmic
age at redshift $z$ and we truncate the formation-redshift integration
at $z_{\rm form,max}=20$. We adopt the fiducial delay-time distribution
$p(t_{\rm d})\propto t_{\rm d}^{-1}$ for
$t_{\rm d}\geq t_{\min}=20\,{\rm Myr}$
\citep{2022ApJ...941..208I}, and the cosmic star-formation-rate density
of \citet{2014ARA&A..52..415M},
\begin{equation}
\dot{\rho}_{\star}(z)
\propto
\frac{(1+z)^{2.7}}
{1+\left[(1+z)/2.9\right]^{5.6}}\,.
\label{eq:SFR}
\end{equation}
Each proposal system is assigned a weight proportional to the geometric multiple-imaging cross-section of the fiducial circular, concentric
NFW+SIS lens \citep{2002ApJ...566..652L}
\begin{equation}
\Omega_{\rm cr}=\pi\beta_{\rm crit}^{2},
\label{eq:cross_section}
\end{equation}
where $\beta_{\rm crit}$ is the source-plane radius of the
multiple-imaging region. Owing to the circular symmetry of the lens,
the source may be placed on the positive $x$ axis. Along the image
branch on the opposite side of the lens, the scalar lens mapping is
\begin{equation}
\beta(\vartheta)
=
\alpha_{\rm NFW}^{\rm Weyl}(\vartheta)
+
\alpha_{\rm SIS}^{\rm Weyl}(\vartheta)
-
\vartheta ,
\label{eq:beta_branch}
\end{equation}
where $\vartheta=|\boldsymbol{\theta}|$ is the angular distance of the image from the lens centre and the $\alpha^{\rm Weyl}$ terms are the radial deflection magnitudes of the two lens components. The critical source radius $\beta_{\rm crit}$ is the largest positive source-plane radius reached by this mapping. The source position is subsequently drawn uniformly in area within the corresponding multiple-imaging region, with $|\boldsymbol{\beta}|=\beta_{\rm crit}\sqrt{u}$, where $u\sim\mathcal{U}(0,1)$, and a polar angle drawn uniformly over $[0,2\pi)$. The final catalogue is drawn from the proposal population with probability proportional to this geometric multiple-imaging cross-section.

For a given event and lens model, we solve the lens equation and calculate the relative arrival times of all recovered images. Our observable is the first-to-last arrival-time span, $\Delta t_{\rm span} = \max_j(t_j)-\min_j(t_j)$. Using this span avoids assigning persistent labels to images whose number or arrival-time ordering may change when the lens model is perturbed. In future analyses of observed systems, the full set of independent time delays between the recovered GW signals can be modelled jointly, retaining more of the information contained in the macroimage arrival times. In this analysis, the mock observation is generated as $
\Delta t_{\rm span}^{\rm obs}
=
\Delta t_{\rm span}^{\rm true}
+
\mathcal{N}(0,\sigma_{\Delta t}^{2})$,
where $
\sigma_{\Delta t}
=
f_{\Delta t}
\left|\Delta t_{\rm span}^{\rm true}\right|$,
and $f_{\Delta t}=0.05$, which is the fiducial
fractional uncertainty on the baseline time-delay prediction. This value defines the reference precision against which the shifts induced by each controlled model mis-specification are assessed. The contribution of the direct GW timing uncertainty to the total time-delay uncertainty is expected to be small, since repeated GW signals can in principle be timed to high precision \citep{2018IAUS..338...98S}. The limiting uncertainty is instead expected to arise from reconstruction of the Fermat-potential difference from the lens--source model. Galaxy-scale forecasts find sub-per-cent statistical precision under an assumed lens model \citep{2017NatCo...8.1148L}, whereas \citet{2019MNRAS.489.2097B} adopt an approximately $5$ per cent level as representative of the total uncertainty in time-delay cosmography arising from the main-deflector model and line-of-sight structure, and identify lens-model complexity as a limiting factor in the cluster regime. We therefore adopt 5 per cent as a fiducial effective uncertainty on the baseline time-delay prediction against which the impact of model mis-specification is assessed. The controlled lens, source and environmental perturbations considered below are introduced separately to determine which forms of model mis-specification would be significant relative to this benchmark. The same noise prescription is applied when the events are analysed separately. For the noiseless closure test, the mock time delays are set equal to their fiducial predictions, without adding a random noise realisation, while the assumed 5 per cent uncertainty is retained in the likelihood. 

We do not explicitly model microlensing in the present procedure \citep[e.g.][]{2026arXiv260617765Z}, as it cannot generally be represented as a simple fractional shift of a macroimage delay because it can introduce frequency-dependent waveform distortions and bias the recovered coalescence time. Its direct contribution to the first-to-last cluster-scale macroimage time-delay span is nevertheless expected to be subdominant to the effective uncertainty adopted here \citep{2026arXiv260916891S}. A waveform-level microlensing injection-and-recovery analysis, particularly for highly magnified events near cluster caustics, is left to future work.

\subsection{Forward model}
\label{subsec:forward_model}

For each mock event, the forward model predicts the time-delay observable as a function of the modified-lensing-potential parameters. For each event $i$ in the baseline case, the input is $\mathcal{E}_i=\left\{z_{l,i},\,z_{s,i},\,M_{l,i},\,c_i,\,\theta_{{\rm E,BCG},i},\,\boldsymbol{\beta}_i\right\}$, where $z_\mathrm{l}$ and $z_\mathrm{s}$ are the lens and source redshifts, $M_l$ and $c$ define the NFW halo profile, $\theta_{\rm E,BCG}$ sets the strength of the central SIS component, and $\boldsymbol{\beta}$ is the unlensed angular position of the source. In the baseline analysis these quantities are fixed to their mock values, while the parameters of the modified Weyl potential are varied. For a given parameter $D$ we construct the total lens model, $\psi_{\rm tot} \left(\boldsymbol{\theta};D,\mathcal{E}_i\right)$. The same Weyl-potential modification is applied to both the cluster-scale NFW halo and the central SIS component. The image positions are obtained by solving the lens equation. Once the image positions have been found, the Fermat potential is evaluated. For event $i$, the lens equation yields a set of image positions $\{\boldsymbol{\theta}_{i,j}\}$, where $j$ labels the individual images. We evaluate the Fermat potential at each image position and
convert it into an arrival time through Eq. \eqref{eq:time_delay} up to a common additive constant. We denote the observable used in this analysis by $\Delta t_{{\rm span},i}^{\rm model}(D)= \max_j\left[t_{i,j}(D)\right]- \min_j\left[t_{i,j}(D)\right]$, which represents the permutation-invariant first-to-last arrival-time span.

The data vector for a catalogue of $N_{\rm ev}$ events is therefore $\mathbf{d}^{\rm model}(D)=\{\Delta t_1^{\rm model}(D),\Delta t_2^{\rm model}(D),\ldots,\Delta t_{N_{\rm ev}}^{\rm model}(D)\}$. Assuming independent Gaussian uncertainties on the measured time delays, the likelihood is
\begin{equation}
\ln \mathcal{L}(D)
=
-\frac{1}{2}
\sum_{i=1}^{N_{\rm ev}}
\left[
\frac{
\Delta t_i^{\rm obs}
-
\Delta t_i^{\rm model}(D)
}{
\sigma_{\Delta t,i}
}
\right]^2
+
{\rm const.}
\label{eq:likelihood}
\end{equation}
where $\Delta t_i^{\rm obs}$ is the mock observed time delay and $\sigma_{\Delta t,i}$ is the corresponding uncertainty. We adopt a uniform prior on $D$ over the interval $-2 \leq D \leq 2\,{\rm kpc}^{-2}$. The frozen event catalogue contains 250 systems and the events are analysed separately in the present study. Once a realistic event forecast and selection model are specified, the individual-event likelihoods could be combined to obtain joint constraints on the modified gravity parameter. 

The lensing calculation is implemented with \texttt{lenstronomy} \citep{2018PDU....22..189B,2021JOSS....6.3283B}. We use custom \texttt{lenstronomy} lens-profile classes to implement the modified NFW and SIS components and solve the lens equation for the image positions. For each recovered image, the relative arrival time is then computed directly from the corresponding Fermat potential, using the lens and source redshifts and the angular-diameter distances in the fiducial background cosmology of \citet[][Table 1]{2020A&A...641A...6P}. The geometric-optics macroimage time delays depend on the lens potential, the lens--source geometry and the associated cosmological distances, but not directly on the intrinsic binary parameters or the detector noise curve. Our results are therefore independent of these quantities because all macroimages returned by the lens equation are included without applying image-level detectability cuts. In an observational analysis, however, the binary masses, spins, inclination and detector sensitivity would affect the signal-to-noise ratios of the individual macroimages, and hence which images are detected, whether an electromagnetic counterpart can be identified, and the precision with which their arrival times are measured.

\begin{figure*}
    \centering
    \includegraphics[width=1\textwidth]{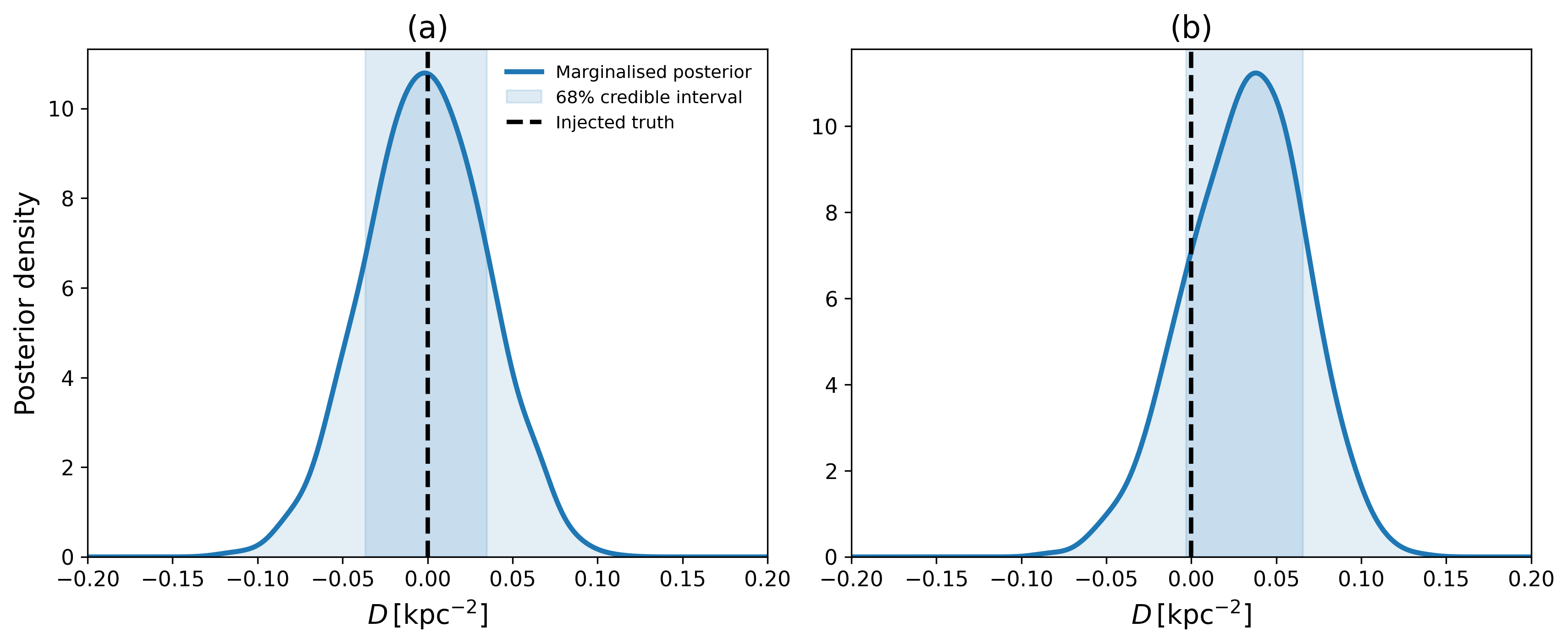}
    \vspace{-7mm}
    \caption{
    Posterior constraints on the modified-gravity amplitude $D$ for the no-systematic case and a lens which produces a minimally asymmetric posterior. (a) Noiseless closure test, in which the mock observable is generated at the fiducial GR value, $D=0$, without adding a noise realisation, while retaining the adopted 5 per cent uncertainty in the likelihood. (b) Noisy baseline, in which Gaussian time-delay noise is added according to the adopted fractional uncertainty. In both panels, the black dashed line marks the injected GR value and the shaded vertical region shows the central 68 per cent credible interval.
    } \label{fig:posterior_no_systematic}
\end{figure*}

To make the posterior exploration computationally feasible, we evaluate the exact forward model on a regular grid in the modified-gravity parameter and interpolate the resulting time delays within the likelihood. The posterior is sampled with \texttt{emcee} \citep{2013PASP..125..306F}, and convergence is assessed using the integrated autocorrelation time together with visual inspection of the ensemble traces. The accuracy of the interpolator is validated with a leave-one-out test, in which grid points are omitted in turn and their time-delay predictions are reconstructed from the remaining points and compared with the direct forward-model evaluations. The interpolation error is small relative to
the adopted time-delay uncertainty throughout the region with appreciable
posterior support. Therefore, the posterior constraints are dominated by the assumed observational errors and by the physical sensitivity of the data vector. We present the closure test for the framework in Fig. \ref{fig:posterior_no_systematic}, which demonstrates that the inference pipeline accurately recovers the injected value of the modified-gravity parameter within the expected posterior uncertainty in the absence of systematic perturbations.

\begin{figure*}
    \centering   \includegraphics[width=1\textwidth]{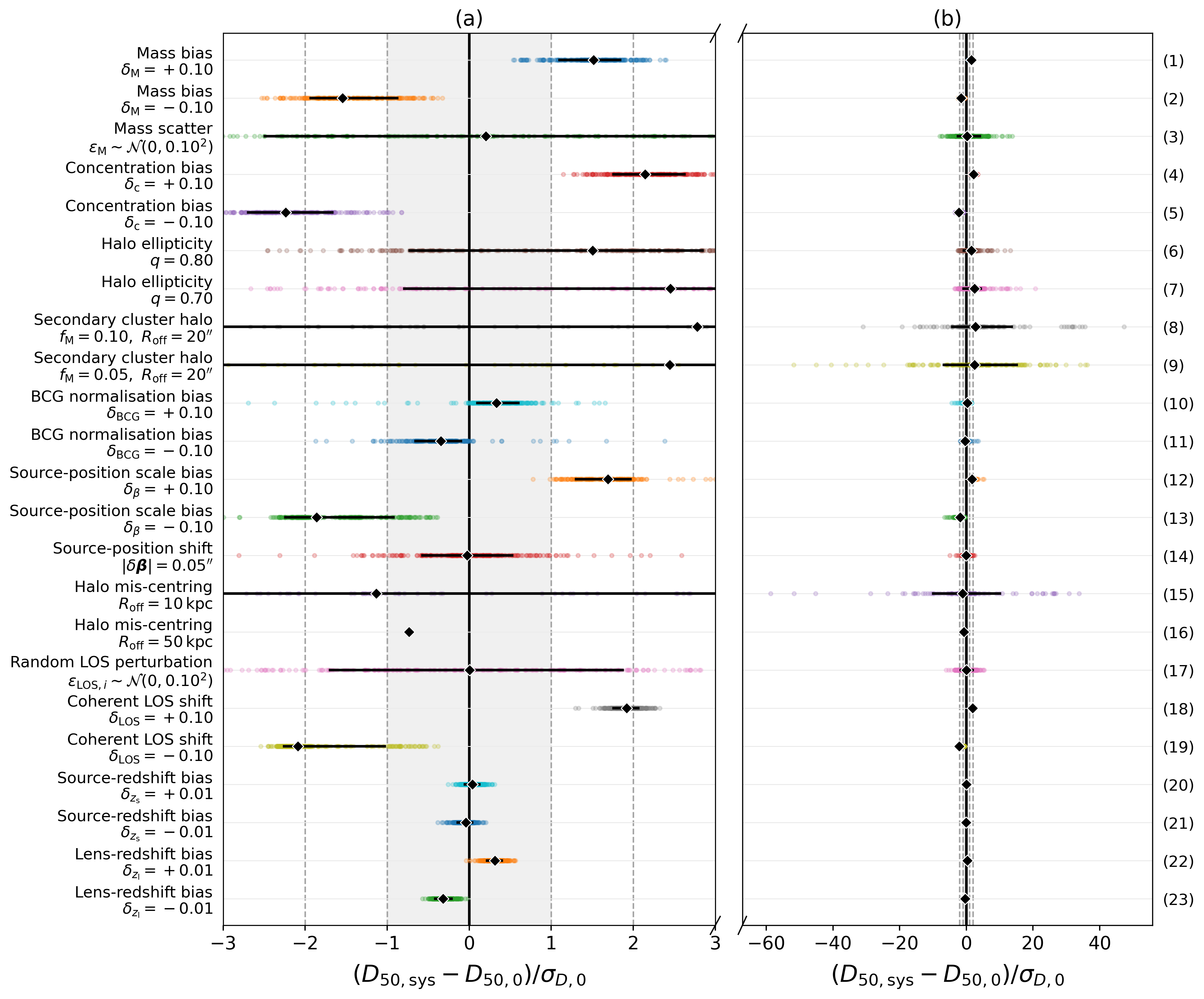}
    \vspace{-6mm}
    \caption{Impact of lens and source modelling systematic effects on the inferred modified-gravity amplitude $D$ for the 250 simulated events. The black diamonds show the median signed baseline-normalised shift,
    $S_s=(D_{50,s}-D_{50,0})/\sigma_{D,0}$, and the horizontal bars show
    the corresponding 16th--84th percentile range across successful
    inferences. The number of successful inferences is case-dependent, since some perturbations can cause the mock configuration to become singly imaged. The exact event counts for each case are reported in Sec. \ref{sec:results}. The coloured circular points represent individual events. Here, $D_{50,s}$ is the posterior median for systematic case $s$, $D_{50,0}$ is the posterior median of the noisy no-systematic baseline for the same event, and $\sigma_{D,0}=(D_{84,0}-D_{16,0})/2$. (a) Zoom around zero over $-3<S_s<3$. (b) Full range of the same results. The solid vertical line marks $S_s=0$, the dashed lines indicate $|S_s|=1$ and $|S_s|=2$, and the shaded region denotes $|S_s|<1$. The left axis identifies each systematic perturbation together with the parameter and injected value used to define it, while the right axis gives the corresponding case number used for reference in the text. Individual points or percentile intervals outside the range of panel (a) remain visible in panel (b). Apparently missing individual points in panel (b) are hidden by overlap with other points or with the median marker.
}
    \label{fig:systematics_summary_D}
\end{figure*}

\subsection{Systematic effects}
\label{subsec:systematics}

We assess the sensitivity of the inferred modified-gravity parameter to controlled mis-specifications of the lens and source model. In each case, the mock observable is generated after perturbing one aspect of the fiducial event, while the inference is performed using the original circular NFW+SIS grid. The injected modified-gravity value is always the GR value, $D=0$. The same random seed and time-delay noise realisation are used in all tests, allowing the resulting posterior shifts to be compared directly with the noisy no-systematic baseline. The amplitudes adopted below span observationally plausible uncertainties. In the ensemble results, coherent biases are identified by a stable signed displacement and a relatively narrow event-to-event spread, whereas configuration-dependent systematic effects produce broad distributions because their impact depends sensitively on the lens--source geometry and may vary in both magnitude and sign. We provide a summary of our results in Fig. \ref{fig:systematics_summary_D}.

\subsubsection{No-systematic baseline}

In the baseline test, the mock time-delay span and the inference model use the same circular Weyl-modified NFW+SIS lens. For the noisy baseline, Gaussian noise is added with standard deviation $\sigma_{\Delta t}=0.05|\Delta t_{\rm span}^{\rm true}|$. For this case, we verify closure with a noiseless test, in which no random noise realisation is added to the mock observable while the likelihood retains the same 5 per cent uncertainty.

\subsubsection{Halo mass}

Uncertainties in cluster mass reconstruction can arise from projection effects, halo triaxiality, correlated structure and assumptions made when fitting the lensing signal. Mock lensing analyses typically find mean or median mass biases of a few to approximately $10$ per cent, with substantially larger object-to-object scatter, often at the
$20$--$30$ per cent level or above \citep{2012MNRAS.421.1073B,2014MNRAS.440.1899G}. We therefore consider a coherent mass bias of $\delta_M=\pm0.10$ and a stochastic scatter of $0.10$ dex, corresponding to an approximately $25$ per cent multiplicative dispersion and characteristic of the high-mass end of mock weak-lensing reconstructions. In the bias test, the mock NFW mass is rescaled as $M_{200c}^{\rm mock}=(1+\delta_M)M_{200c}^{\rm fid}$, while the inference model retains $M_{200c}^{\rm fid}$. In the scatter test, $\log_{10}M_{200c}^{\rm mock}=\log_{10}M_{200c}^{\rm fid}+\epsilon_M$, with $\epsilon_M\sim\mathcal{N}(0,0.10^2)$.

\subsubsection{Halo concentration}

Concentration estimates from spherical NFW fits can be biased by halo orientation, substructure and projected matter along the line of sight \citep{2012MNRAS.421.1073B}. We test this sensitivity by rescaling the mock concentration as $c_{200c}^{\rm mock}=(1+\delta_c)c_{200c}^{\rm fid}$, with $\delta_c=\pm0.10$, while keeping the fiducial concentration fixed during inference. This amplitude is representative of the coherent biases found in mock weak-lensing reconstructions. Mass and concentration are perturbed separately here to isolate their individual effects, although they are generally correlated in observational lens modelling.

\subsubsection{Brightest cluster galaxy}

The central lensing potential receives contributions from both the cluster-scale dark-matter halo and the BCG. In a real lens reconstruction, the BCG contribution must be inferred from strong-lensing constraints, stellar light and stellar-kinematic measurements. These observables help separate the galaxy-scale mass component from the inner cluster halo, but the decomposition remains sensitive to uncertainties in the adopted BCG model, including the density profile, stellar mass-to-light ratio and orbital anisotropy \citep{2015MNRAS.447.1224M,2019A&A...631A.130B}.

To isolate the impact of a coherent error in the strength of the BCG component, we generate the mock observable after rescaling the SIS Einstein radius in Eq. \eqref{eq:SIS} according to $\theta_{\rm E,BCG}^{\rm mock}=(1+\delta_{\rm BCG})\theta_{\rm E,BCG}^{\rm fid}$, while retaining the fiducial value during inference. We adopt $\delta_{\rm BCG}=\pm0.10$ as a representative coherent modelling perturbation. Since $\theta_{\rm E}\propto\sigma_\mathrm{v}^2$ for an SIS, these cases correspond to changes of approximately $\pm 5$ per cent in the equivalent velocity-dispersion parameter, comparable to the level of velocity-dispersion uncertainties assumed in strong-lensing analyses \citep{2019NatSR...911608C}. While the Einstein radius itself can be measured at approximately the $1$ per cent level from high-quality imaging, our perturbation is intended to represent a systematic mis-modelling of the BCG contribution rather than the statistical uncertainty of the Einstein radius measurement.

\subsubsection{Source and lens positions}

Different plausible cluster mass models can produce non-negligible image-plane astrometric residuals \citep{2017MNRAS.470.1809A}. However, a given image-plane residual does not correspond to a unique source-position error, since the mapping between image and source displacements depends on the local lensing Jacobian and therefore on the particular lens--source configuration. Consequently, there is no universal observational uncertainty that can be assigned to the source position. We therefore adopt two phenomenological perturbations designed to probe the sensitivity of the inferred modified-gravity parameter to an incorrect source position.

First, we apply a radial rescaling, $ \boldsymbol{\beta}^{\rm mock} = (1+\delta_\mathrm{\beta})\boldsymbol{\beta}^{\rm fid}$, with $\delta_\mathrm{\beta}=\pm0.10$, which represents a fractional perturbation of the source's radial distance from the lens centre. Its absolute size therefore depends on the fiducial source position, providing a relative measure of source-position uncertainty that is independent of the particular lens configuration. Second, we apply a fixed source-plane displacement, $\boldsymbol{\beta}^{\rm mock}=\boldsymbol{\beta}^{\rm fid}+\delta\boldsymbol{\beta},$ where $|\delta\boldsymbol{\beta}|=0.05^{\prime\prime}$, with a direction fixed by the random seed. This corresponds to a source-plane offset of $0.05^{\prime\prime}$, comparable to the tens-of-milliarcsecond source-plane precision discussed for high-accuracy cluster-lens time-delay cosmography \citep{2019MNRAS.489.2097B}, while remaining substantially smaller than the image-plane residuals typically achieved by current cluster lens models. Together, the fixed and fractional perturbations provide complementary absolute and relative tests of the sensitivity of the inferred modified-gravity parameter to source-position errors. Source-plane uncertainties are configuration-dependent and cannot be represented by a single universal value.

Finally, we test the impact of cluster mis-centring by assuming that the observer incorrectly identifies the BCG position with the centre of the cluster-scale dark-matter halo. We place the BCG at the coordinate origin while displacing the NFW halo centre by $\boldsymbol{\theta}_{\rm off}$, such that the mock lensing potential is $\psi^{\rm mock}(\boldsymbol{\theta})=\psi_{\rm NFW}(\boldsymbol{\theta}+\boldsymbol{\theta}_{\rm off}) + \psi_{\rm SIS}(\boldsymbol{\theta})$. We consider physical offsets of $R_{\mathrm{off}}=10\,{\rm kpc}$ and $R_{\mathrm{off}}=50\,{\rm kpc}$, representative of plausible levels of mis-centring in galaxy clusters \citep{2014MNRAS.439....2V,2016MNRAS.456.2566C}.

\subsubsection{Line-of-sight structure}

Matter outside the principal lens plane can modify the image configuration and Fermat-potential differences through additional convergence, shear and higher-order multi-plane effects. For galaxy-scale time-delay lenses, external convergence has been estimated to contribute up to approximately $5$ per cent to the uncertainty budget \citep{2017MNRAS.467.4220R}. Simulations further show that idealised uniform-convergence configurations can produce shifts of order $10$ per cent \citep{2021MNRAS.504.2224L}.

In the present sensitivity study, we introduce two phenomenological perturbations directly at the level of the time-delay observable. For the random line-of-sight test, the mock
time delay is $\Delta t_i^{\rm mock}=\Delta t_i^{\rm fid}+\epsilon_{{\rm LOS},i}|\Delta t_i^{\rm fid}|$, where $\epsilon_{{\rm LOS},i}\sim\mathcal{N}(0,0.10^2)$. For the coherent shift test, the mock delay is rescaled as $\Delta t_i^{\rm mock}=(1+\delta_{\rm LOS})\Delta t_i^{\rm fid}$, with $\delta_{\rm LOS}=\pm0.10$. For a uniform external convergence, such a rescaling is approximately equivalent in magnitude to $|\kappa_{\rm ext}|\simeq0.10$. This lies near the high-convergence regime explored by \citet{2021MNRAS.504.2224L}. These observable-level perturbations do not reproduce the correlated changes in image positions, magnifications and time delays generated by a physical multi-plane mass distribution. They are intended only to quantify the sensitivity of the inferred modified-gravity parameter to line-of-sight-induced changes of the time-delay scale.

\subsubsection{Redshift uncertainties}

We perturb the source and lens redshifts separately as $z_\mathrm{s}^{\rm mock}=(1+\delta_{z_\mathrm{s}})z_\mathrm{s}^{\rm fid}$ and $z_\mathrm{l}^{\rm mock}=(1+\delta_{z_\mathrm{l}})z_\mathrm{l}^{\rm fid}$, with $\delta_{z_\mathrm{s}}=\delta_{z_\mathrm{l}}=\pm0.01$. The inference model retains the
fiducial redshifts. A 1 per cent source-redshift perturbation is representative of an
optimistic photometric-redshift uncertainty. For the lens redshift, the same amplitude provides a conservative photometric-redshift stress test, since the mean redshift of an identified cluster with multiple spectroscopic members would generally be known considerably more accurately.

\subsubsection{Halo ellipticity}

Observed and simulated galaxy clusters are generally non-circular in projection. For example, from two-dimensional weak-lensing reconstructions of 20 massive CLASH clusters, \citet{2018ApJ...860..104U} measure a median projected minor-to-major axis ratio of $q=0.67\pm0.07$. Triaxial halo shapes are also a generic prediction of collisionless structure-formation simulations, with the axis-ratio distribution depending on halo mass and redshift \citep{2015MNRAS.449.3171B}.

We generate mock data using an elliptical Weyl-modified NFW halo with $q=0.80$ and $q=0.70$, where $q$ denotes the projected minor-to-major axis ratio. Ellipticity is introduced in the projected NFW density through the area-preserving elliptical radius $R_{\rm ell}^{2} = q\,x'^{2} + y'^{2}/q$,  where $(x',y')$ are coordinates aligned with the principal axes of the halo, such that $q=1$ recovers the circular profile. The halo position angle is drawn uniformly over $[0,\pi)$. The value $q=0.70$ is close to the median projected shape measured for the CLASH cluster sample, while $q=0.80$ represents a milder departure from circular symmetry. Together, the two cases probe the response to moderate and representative cluster ellipticities. In both cases, inference is performed using the circular baseline halo with $q=1$, thereby isolating the bias caused by neglecting projected halo ellipticity.

\subsubsection{Secondary halo}

Cosmological ray-tracing simulations show that secondary haloes can alter strong-lensing cross-sections and increase the number of detectable cluster lenses \citep{2019ApJ...878..122L}. Here, we isolate the effect of an unmodelled companion in the main lens plane rather than constructing a full multi-plane matter distribution. We add a second Weyl-modified NFW halo at a projected angular separation of $R_{\rm off}=20^{\prime\prime}$ from the primary halo, with a position angle fixed by the random
seed. Its mass is set to $M_{\rm secondary} = f_M M_{\rm primary}$, with $f_M=0.10$ and $0.05$, corresponding to primary-to-secondary mass ratios of $10{:}1$ and $20{:}1$, respectively, as representative values spanning moderately and strongly perturbed cluster configurations, consistent with the range of multi-component morphologies observed in merging galaxy clusters and used in realistic cluster lens models \citep{2019ApJ...878..122L,2025PhRvD.112f3044V}. The $f_M=0.05$ case represents a relatively subdominant companion, whereas $f_M=0.10$ corresponds to a more strongly perturbed multi-component cluster configuration \citep{2007MNRAS.380..911S}. The secondary-halo concentration is assigned using the same concentration--mass relation as for the primary. Inference is performed using only the original primary NFW halo and BCG, so that any shift in the recovered modified-gravity parameter quantifies the impact of omitting the companion mass component. The fixed $20^{\prime\prime}$ offset is chosen to place the secondary halo near the central strong-lensing region.

For circular configurations, image multiplicity is determined from the radial lens mapping, whereas the asymmetric cases with halo ellipticity and an offset secondary halo require solving the full two-dimensional lens equation. If fewer than two images are recovered, we independently inspect the caustic structure to distinguish genuine single-image configurations from numerical failures. Only configurations confirmed to lie outside the multiple-imaging region are classified as selection losses.

\section{Results}
\label{sec:results}

Fig. \ref{fig:systematics_summary_D} summarises the effect of lens and source modelling systematic effects on the inferred modified-gravity amplitude $D$ for the 250 simulated events, assuming an effective fractional time-delay uncertainty of 5 per cent. For each event and systematic variant, we define the signed normalised shift $S_s = (D_{50,s}-D_{50,0})/\sigma_{D,0}$ and $\sigma_{D,0}=(D_{84,0}-D_{16,0})/2$, where the subscript 0 denotes the noisy no-systematic inference for the same event. The coloured points show individual event-level shifts, while the black diamonds and horizontal bars show the median and 16th–84th percentile range across inferred events. The vertical solid line marks zero shift and the dashed lines indicate $|S_s|=1$ and $|S_s|=2$. If a systematic perturbation changes the lens configuration from multiply imaged to physically singly imaged, the event is classified as a selection loss and no time-delay inference is performed for that event--systematic pair. Such events therefore do not enter the corresponding $S_s$ distribution. Consequently, the number of inferred events can differ between systematic variants.

The normalisation by the event-specific baseline uncertainty allows the systematic displacement to be compared with the statistical precision of each event. A signed ensemble median close to zero does not imply a negligible systematic: stochastic perturbations may induce large event-level shifts of both signs that cancel in the median. For each perturbation, we report the median baseline-normalised shift $S_s$ across the event ensemble together with the 16th--84th percentile interval of the resulting distribution, which characterises the event-to-event spread rather than the uncertainty on the median. In what follows, systematic variants are referred to using the labels shown on the right-hand vertical axis of Fig. \ref{fig:systematics_summary_D}, to make clear which specific perturbation is being discussed. We report the number of multiple-imaging losses for each variant at the end of this section.

\subsection{Lens mass and concentration}
\label{subsubsec:results_mass_concentration}

A coherent $+10$ per cent bias in the primary-halo mass (variant 1) produces a median baseline-normalised shift $S_s=+1.52$, with a 16th--84th percentile range $[+1.10,+1.84]$. The corresponding $-10$ per cent perturbation (variant 2) gives $S_s=-1.54$ and $[-1.93,-0.88]$. The approximately antisymmetric response shows that a coherent mass-normalisation error is preferentially absorbed by $D$, producing a typical displacement larger than the baseline statistical uncertainty. The 0.10-dex mass-scatter test (variant 3) behaves differently. Its signed median is consistent with zero, $S_s=+0.20$, but its 16th--84th percentile range is $[-2.49,+4.04]$. Mass scatter is therefore not negligible: the direction of the induced shift varies among events, causing substantial cancellation in the signed ensemble median while leaving large event-by-event biases. The inference is more sensitive to coherent concentration errors. A $+10$ per cent concentration bias (variant 4) gives a median $S_s=+2.15$ and range $[+1.76,+2.62]$, whereas a $-10$ per cent bias (variant 5) gives $S_s=-2.24$ and $[-2.70,-1.67]$. The central 68 per cent of the inferred event ensemble therefore retains the sign of the injected perturbation, with typical shifts of approximately twice the baseline uncertainty.

This behaviour is physically consistent with the response of our modified lensing kernel for $A=1\,{\rm kpc}^{-2}$ and $R_{\rm scr}=1\,{\rm kpc}$: at the cluster strong-lensing radii relevant here, positive $D$ enhances the effective lensing strength, so underestimating the halo mass or concentration is preferentially compensated by $D>0$, while overestimating them drives $D<0$. For the zero-centred mass-scatter test, these opposite shifts largely cancel in the ensemble median while retaining a broad event-to-event spread. The stronger response to concentration is also physically expected for an NFW halo: at fixed $M_{200c}$, increasing $c$ decreases the scale radius and raises the characteristic inner lensing strength, thereby redistributing more of the halo mass toward the strong-lensing region and producing a larger response than a comparable fractional change in the total halo mass.

\subsection{Halo ellipticity and secondary structure}
\label{subsubsec:results_halo_structure}

Fitting elliptical-halo mocks with the circular baseline model gives a strongly configuration-dependent response. For an axis ratio $q=0.80$ (variant 6), the median shift is $S_s=+1.50$, with a 16th--84th percentile range $[-0.73,+2.85]$. Increasing the ellipticity to $q=0.70$ (variant 7) raises these values to $S_s=+2.45$ and $[-0.79,+4.33]$. The broad ranges, including shifts of both signs, show that the response depends on the halo orientation and source--image configuration rather than only on the axis ratio.

For a secondary-to-primary mass fraction $f_M=0.10$ at an offset of $20^{\prime\prime}$ (variant 8), the median is $S_s=+2.78$, with a 16th--84th percentile range $[-4.19,+13.57]$. For $f_M=0.05$ (variant 9), we find a median $S_s=+2.44$ with range $[-6.72,+15.11]$. Secondary structure is therefore highly consequential in two distinct ways: it can remove a substantial fraction of systems from the multiply-imaged sample, and among the systems that remain, it can generate very large shifts of either sign. The broad distributions indicate a strongly configuration-dependent response rather than a simple monotonic dependence on the secondary-halo mass fraction.

For an elliptical halo, the lensing potential becomes angle-dependent, with the amplitude of the angular perturbation increasing as $q$ decreases. The resulting time-delay shifts therefore vary across events, and larger ellipticity broadens their distribution. An offset secondary halo is likewise intrinsically configuration dependent, since it perturbs different images by different amounts and can modify the critical-curve and caustic structure. The response need not vary monotonically with secondary-halo mass fraction because the reported $S_s$ distributions are conditional on systems remaining multiply imaged, and the two perturbations therefore retain different subsets of the fiducial catalogue.

\subsection{Brightest cluster galaxy}
\label{subsubsec:results_bcg}

A $+10$ per cent bias in the BCG normalisation (variant 10) gives a median $S_s=+0.33$ and a 16th--84th percentile range $[+0.10,+0.60]$. The $-10$ per cent perturbation (variant 11) gives $S_s=-0.34$ and $[-0.66,-0.10]$. The median displacement in $D$ is substantially smaller than that produced by coherent primary-halo mass or concentration errors. Increasing the SIS normalisation strengthens the central potential and is therefore compensated by $D>0$ when the BCG contribution is underestimated, with the opposite shift for an overestimate. Its smaller impact compared with the primary halo is expected because the BCG affects only the innermost part of the lensing potential.

\subsection{Source and lens positions and line-of-sight structure}
\label{subsubsec:results_source_los}

The multiplicative source-position perturbations produce an approximately antisymmetric median response. A $+10$ per cent radial rescaling (variant 12) gives $S_s=+1.69$, with a 16th--84th percentile range $[+1.30,+1.97]$, whereas the $-10$ per cent case (variant 13) gives $S_s=-1.86$ and $[-2.24,-0.93]$. The additive $0.05^{\prime\prime}$ source-plane displacement (variant 14) has a signed median close to zero, $S_s=-0.03$, and a 16th--84th percentile range $[-0.57,+0.52]$. Its effect therefore varies in sign and cannot be assessed from the signed median alone. More generally, the broad tails of the source-position tests reflect the non-linear, configuration-dependent mapping between the source position, image configuration and first-to-last time-delay span. The random 10 per cent line-of-sight perturbation (variant 17) similarly has a signed median consistent with zero, $S_s\simeq 0.00$, and a 16th--84th percentile range $[-1.70,+1.87]$. Random line-of-sight structure can therefore produce relevant event-level shifts even when its conditional-ensemble median cancels. By contrast, the coherent line-of-sight perturbations yield a nearly antisymmetric and much narrower response. The $+10$ per cent case (variant 18) gives $S_s=+1.92$ and $[+1.76,+2.06]$, while the $-10$ per cent case (variant 19) gives $S_s=-2.09$ and $[-2.26,-1.03]$. These perturbations are applied directly to the time-delay observable and should be interpreted as phenomenological proxies rather than as physical multi-plane line-of-sight simulations.

The signs of these responses are consistent with the expected time-delay behaviour. In the circular configurations considered here, moving the source radially farther from the lens centre increases the first-to-last Fermat-potential difference, so an outward source-position bias is compensated by $D>0$, while an inward bias drives $D<0$. A displacement in an arbitrary two-dimensional direction has no preferred sign and therefore largely cancels across the ensemble. Because the coherent line-of-sight perturbations are imposed as a common fractional rescaling of the time-delay span, they project accordingly onto $D$, which approximately rescales the effective Weyl lensing strength in the regime considered here.

A halo mis-centring of $R_{\rm off}=10\,{\rm kpc}$ (variant 15) produces a strongly configuration-dependent response in the inferred modified-gravity parameter, with a median shift $S_s=-1.13$ and a 16th--84th percentile range of $[-9.93,+10.09]$. For the larger offset, $R_{\rm off}=50\,{\rm kpc}$ (variant 16), only one event remains multiply imaged, yielding $S_s=-0.73$. An ensemble-level shift in $D$ is therefore not meaningfully defined for this case. The strong configuration dependence is expected because displacing the cluster-scale halo relative to the BCG changes the deflection field and caustic structure in a manner that depends sensitively on the source position relative to the displaced lens.

\subsection{Redshift uncertainties}
\label{subsubsec:results_redshifts}

A $+1$ per cent source-redshift bias (variant 20) gives a median $S_s=+0.04$ and a 16th--84th percentile range $[-0.05,+0.12]$. The corresponding $-1$ per cent case (variant 21) gives $S_s=-0.04$ and $[-0.14,+0.03]$. The lens-redshift response is coherent but remains subdominant. A $+1$ per cent perturbation (variant 22) gives $S_s=+0.31$ and $[+0.22,+0.40]$, while a $-1$ per cent perturbation (variant 23) gives $S_s=-0.32$ and $[-0.41,-0.22]$. At the amplitudes considered here, redshift calibration is therefore subdominant to coherent errors in the primary-halo mass and concentration, multiplicative source-position errors, halo ellipticity and coherent line-of-sight perturbations. The source-redshift response is particularly weak, while the lens-redshift response is comparable in median normalised magnitude to the BCG-normalisation perturbation.

A relatively weak source-redshift response is expected because $z_{\rm s}$ enters the time-delay prefactor through $D_{\rm s}/D_{\rm ls}$ and the reduced lensing strength through the reciprocal ratio $D_{\rm ls}/D_{\rm s}$, leading to a partial cancellation in the composite NFW+SIS lens. For an SIS this cancellation is exact, so the time delay is independent of $z_{\rm s}$ at fixed lens properties and angular source position. By contrast, increasing $z_{\rm l}$ increases the lens comoving distance and therefore the overall time-delay scale, producing the coherent sign of the lens-redshift response.

Physical selection losses, in which a fiducially multiply-imaged system becomes singly imaged after the systematic perturbation, occur for 0, 21, 12, 0, and 41 of the 250 events in variants 1--5, respectively. We find 2, 2, 142, and 88 events in variants 6--9, 0, 28, 48, 0, and 3 events for variants 10--14, 149 and 249 for variants 15--16, 0 for variants 17--20, and 5, 1, and 2 events for variants 21--23. The largest losses therefore arise for the cluster mis-centring perturbations (variants 15 and 16), followed by the secondary-halo perturbations (variants 8 and 9). The asymmetry in the number of physical selection losses between perturbations of opposite sign reflects the fact that the catalogue is selected to be multiply imaged in the fiducial model: perturbations that weaken the lensing strength or move the source outward can push systems across the multiple-imaging boundary, whereas the opposite perturbations generally move them further inside it.

The inferred magnitude of the systematic shifts is conditional on fixing $A$ and $R_\mathrm{scr}$ in the modified-gravity model. The present tests therefore quantify how modelling errors project onto $D$ within this fixed parametrisation. Allowing these parameters to vary could permit part of a lens-model mismatch to be absorbed along degeneracy directions in the enlarged modified-gravity parameter space, while changes in $R_\mathrm{scr}$ could additionally alter the radial weighting of the Weyl-potential modification across the image configuration. The conclusion of the present analysis is therefore that several plausible modelling perturbations can generate apparent shifts in $D$ comparable to or larger than the assumed statistical uncertainty, whereas their precise magnitudes and relative ranking remain conditional on the adopted modified-gravity parametrisation. 


\section{Conclusions}
\label{sec:conclusions}

We have presented a proof-of-concept framework for testing phenomenological modifications of gravity on the Weyl potential using time delays from GWs produced by BNS mergers and strongly lensed by galaxy clusters. The purpose of this study was to assess how lens-, source-, and environmental modelling uncertainties can bias modified-gravity inference with galaxy-cluster-lensed GWs, whose relative arrival times can be measured with exceptionally high precision. Using a controlled ensemble of simulated lens--BNS source configurations, we quantified how these modelling uncertainties propagate into the inferred modified-gravity parameter $D$.

Our findings, conditional on the adopted gravity model following \citet{2019ApJ...880...50Y}, indicate that neglecting lens ellipticity, secondary lens structure, line-of-sight perturbations and cluster mis-centring can induce significant, configuration-dependent biases in $D$. The latter result suggests that highly dynamically perturbed clusters should be avoided for this analysis. Further, we find that biased estimates of the lens mass and concentration, as well as systematic errors in the source position at the level considered here, can introduce significant coherent biases in $D$. Finally, mis-modelling of the BCG velocity dispersion and of the lens and source redshifts is comparatively less important. These results show how systematic effects considered in strong-lens modelling can mimic departures from GR.

These findings are consistent with recent studies showing that realistic galaxy-cluster structure substantially modifies strong-lensing observables. In particular, \citet{2025PhRvD.112f3044V} find that cluster substructure broadens the distributions of time delays and magnifications and alters image multiplicities and strong-lensing rates, while \citet{2026arXiv260530433R} demonstrate that line-of-sight structure can significantly perturb image configurations and critical-curve morphology in hydrodynamical cluster light cones. Our analysis complements these studies by propagating individual cluster-lens, source, and environmental mis-specifications directly into modified-gravity inference. Recent forecasts such as \citet{2026arXiv260309340M} demonstrate the potential of strongly lensed GW time delays and magnifications for tests of screened gravity; our results show how such constraints can be biased when the assumed lens model differs from the underlying system.

The appropriate treatment of these uncertainties for the inference of modifications to gravity from cluster-lensed GWs should ideally follow a Bayesian forward-modelling approach \citep[e.g.][]{2007NJPh....9..447J}. Specifically, the parameters of sufficiently flexible lens models should be jointly inferred with $D$. Line-of-sight structure requires additional information, ranging from external-convergence or shear terms \citep{2017MNRAS.468.2590S} to multi-plane modelling \citep{2014arXiv1409.0015S,2014MNRAS.443.3631M}. The forward model should also include uncertainties in the BCG, including its velocity dispersion and position relative to the centre of the cluster. Since $D$ is inferred from multiple images, proposed lens--source configurations which lead to single imaging should be rejected and consistently accounted for in the sample selection. Finally, it is interesting to note that source-position uncertainties which should be inferred jointly with $D$ can be reduced in the case of strongly-lensed GWs \citep{2020MNRAS.497..204Y,2025ApJ...993L..57C}.

Given these results, realistic hydrodynamical simulations of cluster lenses will be valuable for characterising lensing systematics at the population level. A full forecast of modified-gravity constraints will additionally require modelling the observable lensed-GW population and associated selection function, including which binaries and macroimages are detectable. With these ingredients, frameworks such as the one developed here can be extended to forecast modified-gravity constraints from ensembles of lensed GW events, jointly exploiting multiple independent time delays within each system. These constraints could further be combined with complementary lensing information encoded in the relative magnifications and lensing-induced phase shifts of the GW images. Finally, on the theory side, it will be equally important to predict how modified gravity modifies the Weyl potential of realistic cluster environments, including ellipticity, baryonic structure, subhaloes and line-of-sight matter. Progress in this direction will require a genuine synthesis of theory, observations and numerical simulations, with lensed GWs offering a natural testbed for bringing these strands together.

\section*{Acknowledgements}\label{acknowledgements}

ET thanks Alan Heavens for helpful discussions. This research made use of ChatGPT (OpenAI; GPT-5.6 Sol) for assistance with language editing and coding. All AI-assisted material was reviewed, tested, and validated by the authors, who take full responsibility for the scientific content. This analysis made use of \texttt{lenstronomy}, a multi-purpose gravitational lens modeling software package \citep{2018PDU....22..189B,2021JOSS....6.3283B}. This work was supported by STFC through Imperial College Astrophysics Consolidated Grant ST/W000989/1. This work made use of the marvin computing server funded by ANR grant ProGraceRay (ANR-23-CE31-0010) and hosted at LUX, Paris Observatory-PSL.

\section*{Data availability}

Data products underlying this article can be made available upon reasonable request to the corresponding author. 



\bibliographystyle{mnras}
\bibliography{main} 



\bsp	
\label{lastpage}
\end{document}